\documentclass[conference]{IEEEtran}
\IEEEoverridecommandlockouts

\usepackage{cite}
\usepackage{amsmath,amssymb,amsfonts}
\usepackage{algorithmic}
\usepackage{graphicx}
\usepackage{textcomp}
\usepackage{xcolor}
\usepackage{kotex}
\usepackage{hyperref}
\usepackage{booktabs}
\usepackage{pifont}
\usepackage{multirow}
\newcommand{\cmark}{\ding{51}}%
\newcommand{\xmark}{\ding{55}}%

\def\BibTeX{{\rm B\kern-.05em{\sc i\kern-.025em b}\kern-.08em
    T\kern-.1667em\lower.7ex\hbox{E}\kern-.125emX}}
\begin{document}

\title{NanoVoice: Efficient Speaker-Adaptive Text-to-Speech for Multiple Speakers \\
 \thanks{* Corresponding Author}
\thanks{This work was supported by Samsung Electronics Co., Ltd (IO231120-07949-01) and Korean Government (2022R1A3B1077720, 2022R1A5A708390811, RS-2022-II220959, BK21 Four Program, IITP-2024-RS-2024-00397085 \& RS-2021-II211343: AI Graduate School Program)}
}

\author{\IEEEauthorblockN{Nohil Park} 
\IEEEauthorblockA{\textit{Electrical and Computer Engineering} \\
\textit{Seoul National University}\\
Seoul, Republic of Korea \\
pnoil2588@snu.ac.kr}
\and
\IEEEauthorblockN{Heeseung Kim} 
\IEEEauthorblockA{\textit{Electrical and Computer Engineering} \\
\textit{Seoul National University}\\
Seoul, Republic of Korea \\
gmltmd789@snu.ac.kr}
\and
\IEEEauthorblockN{Che Hyun Lee} 
\IEEEauthorblockA{\textit{Electrical and Computer Engineering} \\
\textit{Seoul National University}\\
Seoul, Republic of Korea \\
saga1214@snu.ac.kr}
\and
\IEEEauthorblockN{Jooyoung Choi}
\IEEEauthorblockA{\textit{Electrical and Computer Engineering} \\
\textit{Seoul National University}\\
Seoul, Republic of Korea \\
jy\_choi@snu.ac.kr}
\and
\IEEEauthorblockN{Jiheum Yeom}
\IEEEauthorblockA{\textit{Electrical and Computer Engineering} \\
\textit{Seoul National University}\\
Seoul, Republic of Korea \\
quilava1234@snu.ac.kr}
\and
\IEEEauthorblockN{Sungroh Yoon$^{*}$}
\IEEEauthorblockA{\textit{ECE, AIIS, ASRI, INMC, ISRC, and IPAI} \\
\textit{Seoul National University}\\
Seoul, Republic of Korea \\
sryoon@snu.ac.kr}
}

\maketitle

\begin{abstract}
We present NanoVoice, a personalized text-to-speech model that efficiently constructs voice adapters for multiple speakers simultaneously. NanoVoice introduces a batch-wise speaker adaptation technique capable of fine-tuning multiple references in parallel, significantly reducing training time. Beyond building separate adapters for each speaker, we also propose a parameter sharing technique that reduces the number of parameters used for speaker adaptation. By incorporating a novel trainable scale matrix, NanoVoice mitigates potential performance degradation during parameter sharing. NanoVoice achieves performance comparable to the baselines, while training 4 times faster and using 45 percent fewer parameters for speaker adaptation with 40 reference voices. Extensive ablation studies and analysis further validate the efficiency of our model.

\end{abstract}


\section{Introduction}

With the advancement of text-to-speech (TTS) method \cite{Grad-TTS, VITS}, various speaker-adaptive TTS models \cite{VALL-E, Voicebox, NaturalSpeech2} have been introduced to accurately mimic the target speaker's voice. Speaker-adaptive TTS methods are primarily categorized into zero-shot and one-shot approaches. The zero-shot approach \cite{VALL-E, Voicebox, NaturalSpeech2, P-Flow, NaturalSpeech3}, which does not incur additional training costs for adaptation, necessitates a large dataset and numerous parameters to construct a TTS model and often struggles with unique out-of-distribution (OoD) voices. In contrast, the one-shot approach \cite{AdaSpeech2, NeuralVoiceCloning, Moss2020BOFFINTF, hsieh23_interspeech, 10508477} necessitates fine-tuning of a pre-trained multi-speaker TTS model but effectively adapts to the desired speaker’s voice. This fine-tuning based approach not only enhances robustness against OoD data but also reduces the data and model size requirements during the pre-training phase.

Recently, leveraging the capabilities of diffusion-based generative models, various diffusion-based personalization models have been proposed across diverse applications such as text-to-image \cite{DreamBooth}, demonstrating the ability to achieve personalization with minimal data. This trend extends to one-shot TTS \cite{Guided-TTS2, UnitSpeech} as well, where adaptation is enabled by fine-tuning the entire parameters of pre-trained diffusion-based TTS models with just 5-10 seconds of target reference speech. More recent efforts have incorporated parameter-efficient fine-tuning techniques, such as low-rank adaptation (LoRA) \cite{LoRA}, with one-shot TTS \cite{VoiceTailor}, to efficiently perform speaker adaptation with higher speaker similarity.

Including the aforementioned research on personalization, previous works on fine-tuning pre-trained models have predominantly focused on fine-tuning single tasks \cite{prefix-tuning, llama-adapter} or single reference samples \cite{DreamBooth, Guided-TTS2, UnitSpeech}. Recently, the commercialization of various deep learning models has intensified the need to handle multiple queries in parallel \cite{vLLM, orca}. 
This demand is even more critical for fine-tuning methods, as a naive approach would be to fine-tune for each task sequentially, which is computationally inefficient and memory-intensive. As a result, research into more efficient fine-tuning methodologies has been propelled forward,
where a single fine-tuning process aims to address multiple tasks \cite{wen2024batched} or perform personalization using several reference samples simultaneously \cite{CustomDiffusion, gu2023mixofshow}. 
Although such approaches have improved the efficiency of handling multiple queries, their applications in one-shot TTS remain unexplored.

\begin{figure*}[ht]
    \centering
    \includegraphics[width=0.84\linewidth]{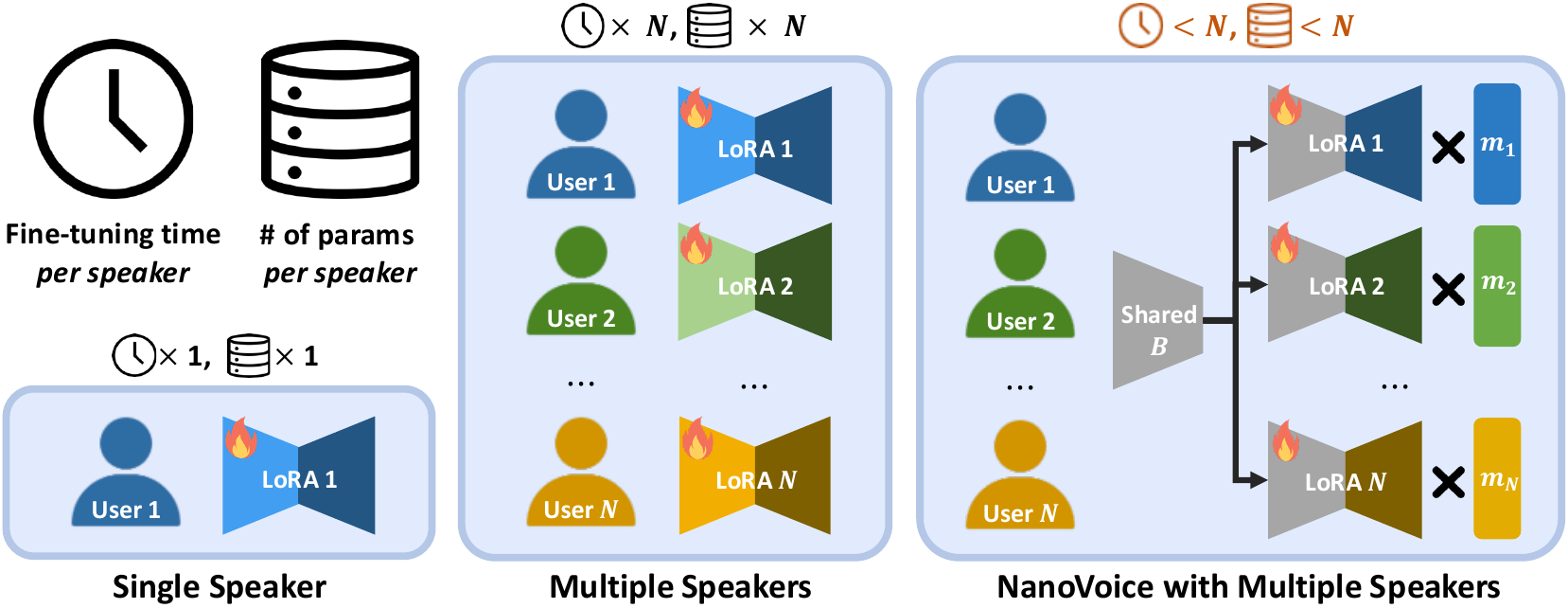}
    \caption{An overview of speaker adaptation in various scenarios: single reference adaptation, sequential adaptation of multiple references, and NanoVoice.}
    \label{fig1} 
    \vskip -0.1in
\end{figure*}

In this work, we propose NanoVoice, an efficient speaker-adaptive TTS designed to perform adaptations for multiple voices simultaneously. Utilizing VoiceTailor \cite{VoiceTailor}, a one-shot TTS that employs LoRA for fine-tuning, as its backbone, NanoVoice accelerates speaker adaptation for multiple references through introducing a batch-wise fine-tuning scheme. Additionally, instead of constructing separate adapters for each of the references, NanoVoice shares parts of the adapters across all references for parameter efficiency. Enhanced by an additional trainable scale matrix, our method enables NanoVoice to adapt to multiple voices efficiently in terms of both parameters and fine-tuning time.

We demonstrate that NanoVoice, when fine-tuned with 40 references for adaptation in parallel, achieves performance comparable to one-shot baselines while being 4 times faster and using 45\% fewer parameters. Moreover, NanoVoice exhibits comparable or superior performance to zero-shot baselines, despite the latter using significantly more data for pre-training. Our ablation studies validate the effectiveness of each component of NanoVoice. Additionally, we conduct several experiments to explore the characteristics and robustness of NanoVoice. Audio samples are on our demo page\footnote{{Demo: \href{https://nanovoice.github.io/}{https://nanovoice.github.io/}}}.

\section{Method}

We introduce NanoVoice, a model that efficiently personalizes multiple reference audios simultaneously. Fig.~\ref{fig1}. compares NanoVoice with sequential fine-tuning. NanoVoice builds upon VoiceTailor \cite{VoiceTailor}, a parameter-efficient one-shot TTS model that advances beyond UnitSpeech \cite{UnitSpeech}, which fine-tunes all parameters, by integrating the LoRA~\cite{LoRA} for parameter-efficient one-shot TTS (Section \ref{method:voicetailor}). NanoVoice improves both time and parameter efficiency via batch operations and parameter sharing (Section \ref{method:batch_wise_finetuning}), with a lightweight scale matrix preventing performance loss (Section \ref{method:lightweight_vecter}).

\subsection{UnitSpeech and VoiceTailor}
\label{method:voicetailor}

UnitSpeech \cite{UnitSpeech}, a diffusion-based \cite{NEURIPS2020_4c5bcfec} one-shot TTS model, first defines the forward process that progressively transforms the mel-spectrogram $X_0$ into noise vector $X_1 \sim N(0, I)$. Given a noise schedule $\beta_t$ and a random noise vector $\epsilon_t \sim N(0, I)$, we can obtain the corrupted mel-spectrogram $X_t$ at any timestep $t$ as follows:
\begin{equation}  
    X_t = \sqrt{\lambda_t} X_0 + \sqrt{1 - \lambda_t} \epsilon_t, \quad \lambda_t = {\rm e}^{-\int_0^t\beta_{s}ds}.
\end{equation}  

To synthesize speech with a voice of a target speaker $S$ and a transcript $c$, the reverse trajectory of the pre-defined forward process is required, and the score $\nabla_{X_t}\log{p(X_t|c,S)}$ for the corrupted mel-spectrogram $X_t$ is needed. Therefore, UnitSpeech trains a network $s_\theta(X_t|c,S)$ to predict this score, which is then used for mel-spectrogram generation. The loss function used for training and the equation for generation involving the score network $s_\theta$ are as follows:
\begin{align}   
    \label{objective}        
        L&(\theta)={\mathbb{E}_{t,X_0,\epsilon_t}[\lVert\sqrt{1 - \lambda_t}s_\theta(X_t|c,S)+\epsilon_t\rVert_2^2]}, \\   
    \label{reverse process}  
        X&_{t-\Delta{t}} = X_t + \beta_t(\frac{1}{2}X_t + s_\theta(X_t|c, S))\Delta{t} + \sqrt{\beta_t\Delta{t}}z_t,
\end{align}
where $z_t$ is a random vector that follows standard normal distribution $N(0, I)$.

UnitSpeech is pre-trained with \eqref{objective} on LibriTTS \cite{zen19_interspeech}, a large-scale multi-speaker TTS dataset to predict scores for multiple speakers following \cite{Grad-TTS}. During fine-tuning, it adapts all of its parameters using the reference data of the target speaker. Unlike UnitSpeech, VoiceTailor \cite{VoiceTailor} achieves parameter-efficient speaker adaptation by injecting a low-rank adapter into the linear layers of the attention modules in the pre-trained TTS model and fine-tuning only the adapter parameters. Specifically, for a linear layer with matrix $W_0$, VoiceTailor injects a matrix $\Delta W = B A$ where $B \in \mathbb{R}^{d \times r}$ and $A \in \mathbb{R}^{r \times k}$ with a scale factor $\alpha$, resulting in $W = W_0+\alpha \cdot B A$. By setting $r$ much smaller than $d$ and $k$, the number of parameters in $B$ and $A$ becomes significantly smaller than in $W_0$, allowing for fine-tuning with far fewer parameters. In this paper, we follow the observation that $r=2$ is sufficient for effective speaker adaptation in \cite{VoiceTailor}, and therefore set $r$ of NanoVoice to 2.

\subsection{Batch-Wise Fine-tuning Scheme with Parameter Sharing}
\label{method:batch_wise_finetuning}

NanoVoice extends VoiceTailor by building multiple adapters simultaneously using multiple reference voices. Given $N$ reference voices, we first batch the $N$ reference speeches to create a batched reference sample $X'_0\in \mathbb{R}^{N \times L}$, where $L$ is the maximum length of the mel-spectrograms of the reference samples. We then stack $N$ low-rank matrices for each reference to form new matrices $B' \in \mathbb{R}^{b \times d \times r}$ and $A' \in \mathbb{R}^{b \times r \times k}$. During fine-tuning, batch-wise matrix multiplication ensures that the loss and gradient for each reference sample are calculated separately. This approach, performing independent computations within the batch, maintains performance comparable to the method of constructing multiple adapters sequentially while achieving faster speaker adaptation.

In pursuit of further efficiency, we propose to share parameters that are less critical for personalization across multiple voices. To investigate which parameters to share, we run an experiment by selecting 50 random samples from the \texttt{test-clean} subset of LibriTTS \cite{zen19_interspeech} as reference data. Using this, we perform speaker adaptation with four configurations of low-rank adapters: the baseline setup with both matrices batch-wise $(B', A')$, a setup where $B$ is shared across all references $(B, A')$, a setup where $A$ is shared across all references $(B', A)$, and a setup where both matrices are shared $(B, A)$. For each, we build personalized speech models and measure the speaker similarity between the generated and the reference speech, as described in Section \ref{experiments:experimental_setup}.

The results show that sharing $B$ led to the least performance drop, with SECS values of $0.938$ for the batch-wise setup, $0.933$ for sharing $B$, $0.902$ for sharing $A$, and $0.898$ for sharing both $B$ and $A$. Therefore, NanoVoice utilizes $A'$ in a batch-wise manner and shares $B$ across all reference voices. Given that the number of parameters for $B'$ constitutes approximately two-thirds of the total trainable parameters, this sharing approach allowed us to reduce the number of parameters by nearly threefold.

\subsection{Lightweight Scale Matrix}
\label{method:lightweight_vecter}

While the proposed method with batch-wise matrix $A'$ combined with a shared matrix $B$ reduces the number of trainable parameters, it introduces a slight performance degradation, as seen in the previous section. To mitigate this, inspired by DoRA \cite{liu2024dora}, one of several methods for boosting the capacity of LoRA with a minimal increase in parameters \cite{liu2024dora, kopiczko2024vera, nikdan2024rosa}, we introduce a trainable scale matrix $m' \in \mathbb{R}^{N \times 1 \times k}$, composed of stacked scale vectors for each reference.

The scale matrix $m'$ is initialized with the column-wise weight norm of the pre-trained model $W_0$. To compute $W$ during fine-tuning and inference, rather than directly applying $m'$ to $W_0 + \alpha \cdot B A'$, we first normalize $W_0 + \alpha \cdot B A'$ by its column-wise norm, denoted as $||W_0 + \alpha \cdot B A'||_c$. Subsequently, we perform an element-wise multiplication with the scale matrix $m'$, similar to \cite{liu2024dora}. Note that $m'$ is batched across multiple speaker references, in contrast to training a single scale vector \cite{liu2024dora}. Although this method involves fewer additional parameters compared to the original low-rank adapter, it effectively enhances performance, making it a parameter-efficient solution for improving speaker adaptation.

\section{Experiments}

\begin{table*}
\caption{The results of 5-scale MOS, CER, 5-scale SMOS for one-/zero-shot TTS models tested on LibriSpeech. \textbf{\# Params} and \textbf{Adaptation Time} refer to the number of trainable parameters per speaker and the adaptation time per speaker, respectively.}
\vskip -0.1in
\label{table:main}
\begin{center}
\begin{tabular}{l|c|c|cc|ccc}
\toprule
\textbf{Method} & \textbf{Amount of Dataset} & \textbf{Fine-tuning}  & \textbf{\# Params} & \textbf{Adaptation Time} & \textbf{5-scale MOS} & \textbf{CER (\%)} & \textbf{5-scale SMOS} \\ \midrule
Ground Truth    & $-$                        & $-$                   & $-$                & $-$                      & $4.20\pm0.08$        &     0.82\%              & $3.89\pm0.07$         \\
 \midrule
NanoVoice       & $\approx 585$ hrs          & \cmark & $21K$              & $7.6s$                   & $4.10\pm0.09$        &      1.10\%              & $3.88\pm0.08$         \\
VoiceTailor \cite{VoiceTailor}     & $\approx 585$ hrs          & \cmark & $39K$              & $31s$                    & $4.01\pm0.10$        &     1.17\%               & $3.84\pm0.09$         \\
UnitSpeech \cite{UnitSpeech}      & $\approx 585$ hrs          & \cmark & $119M$             &      $32s$                     & $4.06\pm0.09$        &         1.14\%          & $3.85\pm0.09$         \\ \midrule
XTTS \textit{v2} \cite{casanova24_interspeech}         & 27,281 hrs                 & \xmark & $0$                & $0s$                        & $4.00\pm0.09$        &          1.26\%          & $3.74\pm0.09$         \\
CosyVoice \cite{CosyVoice}      & 171,800 hrs                & \xmark & $0$                & $0s$                        & $4.14\pm0.09$        &          3.05\%         & $3.86\pm0.08$  \\ \bottomrule
\end{tabular}
\end{center}
\vskip -0.1in
\end{table*}
\begin{table}[htbp]
\vskip -0.2in
\caption{Ablation studies for adapter sharing. \textbf{\# Params} refers to the number of trainable parameters per speaker.}
\vskip -0.1in
\label{table:ablation_share}
\begin{center}
\begin{tabular}{cc|c|cc}
\toprule
\textbf{Share B}   & \textbf{Share A}   & \multicolumn{1}{l|}{\textbf{\# Params}} & \textbf{CER (\%)} & \textbf{SECS} \\ \midrule
\xmark   & \xmark                         & $38,920$                                & 1.26\%            & 0.870         \\
\cmark   & \xmark  & $14,459$                                & 1.23\%            & 0.868         \\
\xmark   & \cmark  & $25,442$                                & 1.18\%            & 0.815         \\
\cmark   & \cmark    & $973$                                   & 1.13\%            & 0.813         \\  \bottomrule
\end{tabular}
\end{center}
\end{table}

\begin{table}
\vskip -0.2in
\caption{Ablation studies for the trainable scale matrix. \textbf{\# Params} refers to the number of trainable parameters per speaker.}
\vskip -0.1in
\label{table:ablation_scale}
\begin{center}
\begin{tabular}{l|c|cc}
\toprule
\textbf{Methods}                                                                  & \multicolumn{1}{l|}{\textbf{\# Params}} & \textbf{CER (\%)} & \textbf{SECS} \\ \midrule
NanoVoice                      & $21,363$                                & 1.10\%            & 0.873         \\      
\textminus~Normalization              & $21,363$                                &       1.47\%            &      0.865         \\
\textminus~Scale Matrix & $14,459$                                & 1.23\%            & 0.868        \\ \bottomrule
\end{tabular}
\end{center}
\vskip -0.3in
\end{table}

\subsection{Experimental Setup}
\label{experiments:experimental_setup}

\textbf{Datasets.} We use the LibriTTS dataset \cite{zen19_interspeech}, which contains 2,456 speakers, to train the pre-trained TTS model, following the same method as in UnitSpeech and VoiceTailor. For fine-tuning and evaluation, we utilize the \texttt{test-clean} subset of LibriSpeech, consisting of 40 speakers (20 male and 20 female). For each speaker, one reference audio is used for fine-tuning and one transcript is used for generation.

\textbf{Pre-training and Fine-tuning Details.} We follow VoiceTailor's procedure for pre-training a multi-speaker TTS model. For fine-tuning, we train the LoRA weights and scale matrices for 500 iterations using the Adam optimizer \cite{Adam} with a learning rate of 0.0001. We set the LoRA rank to 2 and the scaling factor $\alpha$ to 8. This results in 21,363 trainable parameters \textit{per speaker} for NanoVoice. Single-speaker adaptation takes approximately 7.6 seconds on a single NVIDIA A40 GPU.

\textbf{Evaluations.} NanoVoice, a one-shot TTS model capable of fine-tuning multiple references in parallel, uses adaptation models UnitSpeech \cite{UnitSpeech} and VoiceTailor as baselines. We also use zero-shot TTS models, XTTS \textit{v2} \cite{casanova24_interspeech} and CosyVoice \cite{CosyVoice} for comparison. NanoVoice utilizes the official BigVGAN checkpoint \cite{lee2023bigvgan} as its vocoder, with sampling procedure following the method used in VoiceTailor.

We perform both qualitative and quantitative evaluations using 40 sentences from the LibriSpeech \texttt{test-clean} subset. We use MTurk to measure 5-scale mean opinion score (MOS) for evaluating both audio quality and naturalness, as well as 5-scale speaker similarity mean opinion score (SMOS) for assessing speaker similarity. Additionally, speaker encoder cosine similarity (SECS) is measured using Resemblyzer speaker encoder \cite{resemblyzer}, and pronunciation accuracy is assessed with character error rate (CER) using CTC-based Conformer \cite{Conformer} of NEMO toolkit \cite{kuchaiev2019nemo}. All samples are generated five times using different seeds for fair comparison, and the average SECS and CER values are reported.

\subsection{Model Comparison}

As shown in Table \ref{table:main}, NanoVoice shows comparable MOS and SMOS to other one-shot TTS baselines while using fewer trainable parameters \textit{per speaker}. Notably, NanoVoice achieves speaker adaptation with just 21K trainable parameters, which is less than 0.02\% of the full fine-tuning requirements of UnitSpeech and approximately 53.8\% of the LoRA-based adaptation used by VoiceTailor. Additionally, NanoVoice’s capacity to adapt to multiple speakers simultaneously enables a training speed 4 times faster than that of baseline models.

We also compare NanoVoice with the zero-shot TTS models. XTTS \textit{v2}, which uses approximately 50 times the amount of NanoVoice's pre-training data, shows degraded SMOS performance compared to NanoVoice ($p < 0.05$). As demonstrated in the comparison with CosyVoice, zero-shot TTS models require nearly 300 times larger dataset to achieve similar levels of quality, naturalness, and speaker similarity as NanoVoice, which is reflected in the MOS and SMOS scores.

\subsection{Ablation Studies} In Table \ref{table:ablation_share}, the ablations on the LibriSpeech \texttt{test-clean} subset reaffirm the results of the experiments discussed in Section \ref{method:batch_wise_finetuning}. Notably, sharing $B$ across all references shows comparable SECS while using only 37.1\% of the trainable parameters compared to the batch-wise adapter setup, without sharing any matrices. In contrast, sharing $A$ results in a 1.75-fold increase in trainable parameters but leads to a decrease in performance compared to sharing $B$. The most significant performance drop occurs when both adapters are shared, implying that a single LoRA for all speakers is less effective.

In Table \ref{table:ablation_scale}, we present ablation experiments on the trainable scale matrix and operations proposed in Section \ref{method:lightweight_vecter}. The most intuitive approach to improve the training capacity of the shared matrix $B$ with batched $A'$ is to directly multiply $m'$ without normalizing with $||W_0 + \alpha \cdot B A'||_c$. However, the results show that multiplying only the scale matrix slightly degrades SECS. This suggests that NanoVoice can enhance speaker similarity by ensuring training stability through the combined use of the normalization term.

\subsection{Analysis}
\textbf{Number of Speakers.} In our experiments, we train NanoVoice on the LibriSpeech \texttt{test-clean} subset by batching one reference audio per speaker and matching the number of voice adapters to the batch size. To confirm the robustness of NanoVoice's speaker similarity performance, we conduct an experiment with varying batch sizes. Since NanoVoice trains adapter groups for multiple speakers, Table \ref{table:analysis} shows that as the number of speakers trained simultaneously increases, parameter efficiency improves. Moreover, we observe that NanoVoice maintains consistent speaker similarity even when the number of batched adapters varies.

\textbf{Role of Shared Matrix $\mathbf{\textit{B}}$.} An important observation from Section \ref{method:batch_wise_finetuning} and Table \ref{table:ablation_share} is that sharing the matrix $B$ is sufficient for multi-speaker adaptation. We hypothesize that this observation is due to either $B$ effectively models common information within the reference audio group, or $B$ is not critical for speaker adaptation. To test these hypotheses, we conduct two analyses.

First, to examine whether sharing $B$ benefits the model, we analyze its ability to model gender, a distinct feature in speech. Using LibriSpeech's metadata, we create subsets with a batch size of 20, organized into \textit{same}-gender and \textit{mixed}-gender groups. The \textit{same}-gender groups include two batches, either of 20 male or 20 female speakers, while the \textit{mixed}-gender groups contain two batches with 10 male and 10 female speakers each. The results in Table \ref{table:analysis} show that training NanoVoice on each group yields similar performance, indicating that $B$ does not capture common information within references, thus contradicting the first hypothesis.

Next, to test whether $B$ is crucial for speaker adaptation, we freeze $B$ and fine-tune only the remaining trainable parameters of NanoVoice. With the frozen matrix $B$, NanoVoice achieves SECS of 0.871, only 0.002 points lower than the original NanoVoice (0.873). This suggests that matrix $B$ is not essential for multi-speaker adaptation, aligning with recent findings in the NLP field \cite{lora-fa}. Therefore, we conclude that the effectiveness of parameter-efficient sharing in NanoVoice arises from the fact that $B$ is not critical for speaker adaptation.

\begin{table}
\vskip -0.2in
\caption{Analysis on the number of speakers and the role of shared matrix $B$ in terms of gender attribute.}
\vskip -0.1in
\label{table:analysis}
\begin{center}
\begin{tabular}{l|c|c|cc}
\toprule
\textbf{Gender} & \textbf{Batch Size} & \textbf{\# Params} & \textbf{CER} (\%) & \textbf{SECS}    \\ \midrule
$-$    & $1$        & $45,832$  & 1.30\%   & $0.873$ \\
Mix    & $5$        & $25,755$  & 1.07\%   & $0.872$ \\
Mix    & $20$       & $21,991$  & 1.18\%   & $0.872$ \\
Mix    & $40$       & $21,363$  & 1.10\%   & $0.873$ \\ \midrule
Same   & $20$       & $21,991$  & 1.02\%   & $0.872$ \\ \bottomrule
\end{tabular}
\end{center}
\vskip -0.2in
\end{table}

\section{Conclusion}
In this work, we introduced NanoVoice, a parameter-efficient speaker-adaptive TTS method capable of handling multiple reference voices simultaneously. By employing a batch-wise training scheme and parameter sharing, along with learnable scale matrices, we have significantly reduced both the number of parameters and adaptation time per speaker. We hope that NanoVoice will pave the way for new opportunities in the commercialization of personalized TTS systems.

\bibliographystyle{IEEEtran}
\bibliography{main}

\end{document}